# Usability study of tactile and voice interaction modes by people with disabilities for home automation controls

Nadine Vigouroux[1], Frédéric Vella[1], Gaëlle Lepage[2], Eric Campo[3],

[1] IRIT, CNRS 5505, Université Paul Sabatier, 118 Route de Narbonne, 31062 Toulouse, France
[2] GIHP 10 Rue Jean Gilles, 31100 Toulouse / UT2J 5 allée Antonio Machado, 31058 Toulouse, France
[3] LAAS-CNRS, Université de Toulouse, CNRS, UT2J, Toulouse, France
`nadine.vigouroux@irit.fr`

**Abstract.**
This paper presents a comparative usability study on tactile and vocal interaction modes for home automation control of equipment at home for different profiles of disabled people. The study is related to the HIP HOPE project concerning the construction of 19 inclusive housing in the Toulouse metropolitan area in France. The experimentation took place in a living lab with 7 different disabled people who realize realistic use cases. The USE and UEQ questionnaires were selected as usability tools. The first results show that both interfaces are easy to learn but that usefulness and ease of use dimensions need to be improved. This study shows that there is real need for multimodality between touch and voice interaction to control the smart home. This study also shows that there is need to adapt the interface and the environment to the person's disability.

## 1    Introduction

In France, the ELAN law of November 23, 2018[1] introduces the concept of inclusive housing and defines it as housing mode "intended for people with disabilities and the elderly who have the choice, as their main residence, of a grouped mode of living, among themselves or with other people (...) and accompanied by a social and shared life project". Due to the evolutions and changes in behavior that this new type of housing may entail, it seems interesting to study how the technological needs of people with disabilities could them to improve their living conditions in autonomy while taking into account their physical and material environment. Indeed, digital technologies have shown their potential to compensate for certain difficulties encountered in the daily life of people with disabilities in their homes. According to Khomiakoff [1], "assistive technologies can play a particularly important role in the choice to remain at home, or allow, in addition to adequate social support, greater autonomy and a better quality of life. However, the adequacy of the technological solutions developed to the real need of people with disabilities remains a challenge.

---

[1] Excerpt from Article L.281-1 of the CASF

Vacher *et al.* [2] described an audio-based interaction technology that lets the user have full control over her home environment and at detecting distress situations for the elderly and visually impaired people. Varriale *et al.* [3] identified the role and function of home automation, for people with disabilities through a deep review, in particular they aim to highlight if and how home automation solutions can support people with disabilities improve their social inclusion. Cheng *et al.* [4] investigated the effects of button and spacing size on touchscreen performance by people with varying motor abilities. Mtshali and Khubisa [**Erreur ! Source du renvoi introuvable.**] designed a smart home appliance control system for people with physical disabilities based on a voice digital assistant. Noda [5] reported that persons with disability could utilize the voice applications such as Google Home to control appliances in a smart house.

However, too few interfaces and smart home devices are not designed with people with disabilities and people with limited range of motion, sight, hearing or speech difficulties. In the framework of to the HIP HOPE project concerning a building on the Montaudran site in Toulouse (France) in which 19 inclusive housing units will be built, we conduct a pre-study on the tactile and vocal interaction modes for the home automation control of equipment by disabled people for an inclusive housing. In [6] the authors describe the respective rate of use of voice and touch commands and interaction errors due to their impairment.

Firstly, the paper briefly describes the experiment conducted in the Smart Home of Blagnac[2] in France. Then, we present and discuss the results of the USE and UEQ questionnaires.

## 2        Experiment

### 2.1        Material

We carried out the experimentation in the living lab MIB. It is a 70m² apartment allowing to carry out design groups and experiments with end users (disabled and elderly people). It is composed of different rooms: living room, kitchen, corridor, bedroom, bathroom and toilet. It is equipped with various connected objects such as a removable sink and washbasin, lights, shutters, television, a fall sensor and an electric bed. It also has an infrastructure [8] to support communication between connected objects and control device, and microphones, cameras and motion sensors for sensors. Voice and/or touch interaction is used to activate the connected objects. The Amazon Fire Cube TV personal assistant realizes the voice interaction. The touch interface was designed under OpenHab's HABPanel and installed on a Samsung Tab A7 touch tablet. For example, the participants can formulate the command "open the living room shutter", "turn on the bedroom light" or call "help". In the same way, the tactile command, thanks to "presses/clicks" on a tablet allows to realize the same commands.

---

[2] MIB, http://mi.iut-blagnac.fr

## 2.2 Population

7 persons with disabilities (1 mental disability, 4 motor disabilities including 2 with speech disorders, 1 visual and 1 hearing impairments) participated in this study (see Table 1). We recruited participants of all ages and with different impairments. This set of participants represents the population that will live in the HIP HOPE home automation flats. Table 1 also lists the home automation and assistive technologies desired by these individuals, collected from interviews.

**Table 1.** Table of participants.

| Participant | Age/genre | / Impairment | Activities | Technology needs for smart home |
|---|---|---|---|---|
| 101 | 63/M | Hearing impairment | Pharmacist, now retired | adapted intercom with high quality visuals to see the person and read their lips ; connected objects with visual feedback; flashing lights; app on phone to detect someone's presence or an abnormal noise. |
| 102 | 72/M | Visual impairment | Computer science now retired | easy to implement; efficient and responsive technology, limit the number of steps, preference for voice control with voice feedback on actions performed; home automation control (shutters, light, alarm) but with reliability and ease of use. |
| 104 | 39/F | Cerebral palsy | Employee in an association and volunteer | interfaces for home automation control (shutters, front door); voice control difficult in case of fatigue, so have the touch mode; connected intercom without the need to pick up the phone. |
| 202 | 18/M | Trisomy syndrome | Student | smartphone application to help organise activities, to encourage initiatives (coaching application). |
| 204 | 19/M | Cerebral palsy | Student | smartphone control system for gates, garages and front doors to be autonomous; smartphone remote control for TV, robotic arm. |
| 300 | 38/F | Myopathy | Volunteer | home automation to control the environment (with voice command); robotic arm (help for cutting, grabbing objects, grooming), adapted intercom (easy to open and to communicate). |
| 302 | 70/F | Polio | Secretary, retired and volunteer | opening of the gate from your home; automated bay window; automation control of equipment for individual and mobile homes; fall detector or easy emergency call. |

### 2.3　Courses of the experiment

First, we introduced the participants to the use of the two interaction modes (presentation by the experimenter and learning by the participant). Then they were asked to perform two scenarios (one controlled and one free). These scenarios included tasks (opening shutters, turning on lights, etc. see [6] for a detailed description). Subjects were free to use the tactile or voice command in any order they wished. At the end of the experiment, participants were asked to fill out questionnaires.

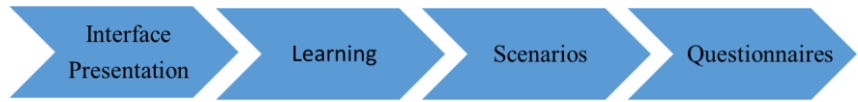

**Fig. 1.** Courses of the experiment.

### 2.4　Questionnaires

We use two questionnaires: USE (Usefulness, Satisfaction, and Ease of use, [9]) and UEQ (User Experience Questionnaire, [10]) and a complementary questionnaire on interaction modalities (modality preference and multimodality).

The USE questionnaire consists of 30 items, divided into 4 dimensions (usefulness, ease of use, ease of learning, and finally satisfaction with its use). Each item is presented in the form of several statements to be noted from 1 to 7. 1 corresponds to "Very disagree" and 7 to "Very agree". The participants filled out the questionnaires just after the scenarios had been run, followed by a debriefing with the experimenters.

The UEQ questionnaire includes 26 items divided into 6 dimensions evaluating the *attractiveness* (general impression), *perspicuity* (easy to get familiar), *efficiency* (fast, efficient, organized), *dependability* (understandable, instinctive), *reliability* (control, predictable), *stimulation* (interest and motivation) and *novelty* (creative, innovative) of the system.

## 3　Results

Both interfaces were evaluated in a general way, without distinction between touch and voice.

### 3.1　Experimental context

Table 2 illustrates the interaction environment (modality, device, tablet placement, mobility support for the participant). Participant 102 used only voice interaction (visual impairment) while participant 204 used only tactile interaction (speech impairment). The other 5 participants used all interaction modalities.

**Table 2.** Interaction environment.

| Participant | 101 | 102 | 104 | 202 | 204 | 300 | 302 |
|---|---|---|---|---|---|---|---|
| Modalities | Touch and voice | Voice | Touch and voice | Touch and voice | Touch | Touch and voice | Touch and voice |
| Interaction devices | | | Help to open the voice channel | | Joystick connected to the tablet | | |
| Position of the tablet | In the hand | No use of the tablet | On the knees | In the hand | On the kitchen table (not mobile) | On the knees | Holds the tablet in one hand and touches with the other hand |
| Movement of the person | Without assistance | Without assistance | In an electric wheelchair | Without assistance | In an electric wheelchair | In an electric wheelchair | In an electric wheelchair |
| Touch interaction | 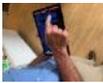 | 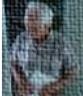 | 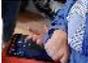 | 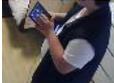 | | 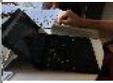 | 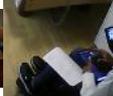 |

### 3.2 USE Questionnaire

The participant filled the questionnaire after the scenarios were played. For the analysis of these data, the 7 items were transposed to scores ranging from -3 (totally disagree) to 3 (totally agree) in order to have more contrasting results.

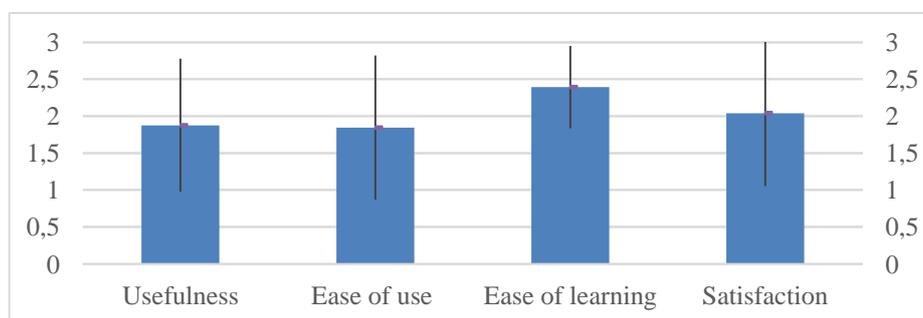

**Fig. 2.** Mean and standard deviation of the USE questionnaire score for the 7 participants.

Figure 2 shows the mean and standard deviation of the four dimensions of the USE questionnaire for 7 participants. The USE rating for *Ease of learning* (2,4) and *Satisfaction* dimension (2) are good, even very good for *Ease of learning*. We can see that Usefulness and Ease of Use rating are similar (1.8). For 3 of the 4 dimensions, we find a significant standard deviation (±-0.9), except for the *Ease of learning* (±0.5).

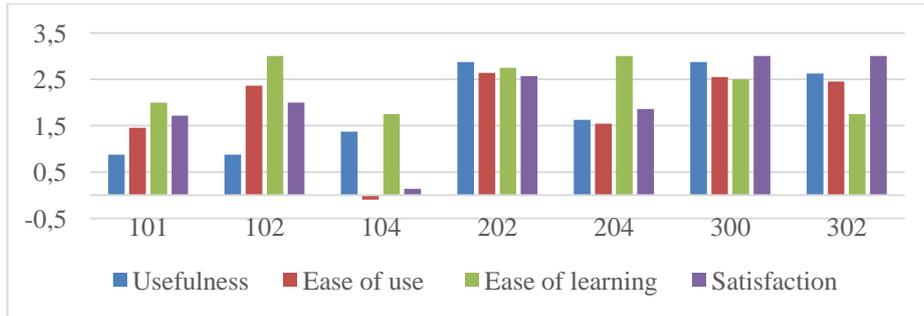

**Fig. 3.** Mean and standard deviation of the USE questionnaire score for the 7 participants.

Figure 3 shows that participants 101, 104 and 204 have some difficulty using the interaction modalities (*Ease of Use*) or even very serious difficulty using them (participant 104). These difficulties in using the interaction modalities resulted in a decrease in *satisfaction*. 4 participants found learning the interaction modalities very easy (>2.5) and 3 easy (>1.5). The Usefulness dimension was low for two participants (<1) and two quite low (>1 and <1.5).

### 3.3 UEQ Questionnaire

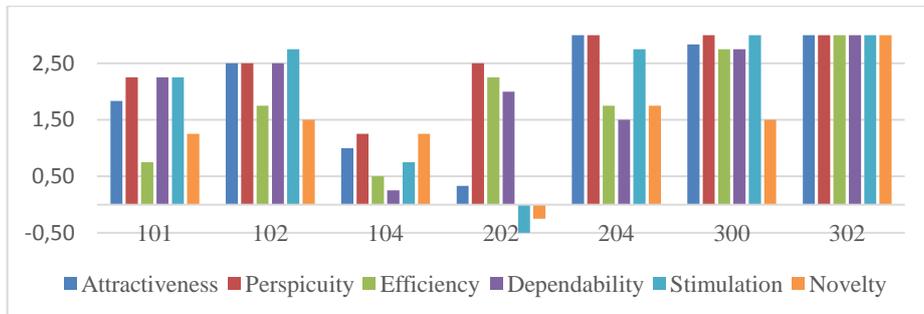

**Fig. 4.** Mean and standard deviation of the UEQ questionnaire score.

The UEQ is scored as the USE questionnaire. The UEQ value between -0.8 and 0.8 represent a more or less neutral evaluation of the corresponding dimension, values > 0.8 represent a positive evaluation and values < -0.8 represent a negative evaluation. UEQ scores (*attractiveness, perspicuity, and stimulation*) are highly positive (>2) for four participants while the criteria *novelty* is lower (>1.5 and <1.8) for five participants as shown in Figure 4. For participant 104, all dimensions of the UEQ question-naire are < 1.25. Negative values for the dimensions stimulation and novelty are also noted for participant 202.

## 4     Discussion

Participants enjoyed participating in the experiment. They discovered the possibilities of voice and touch interaction with the connected objects of the smart home. The least positive scores were related to "ease of use" and "usefulness" of the USE questionnaire. Not all interface accessibility conditions were met for participants 101 (no magnetic loop and no visual feedback), 102 (no screen reader running on the tablet), 104 (severe speech difficulties and therefore recognition problem for the voice assistant). This study clearly shows that these interfaces alone are not sufficient and that the experimental environment must also incorporate assistive technologies for the hearing and vision impaired. The variability in *usefulness* dimension may be due to the limited exposure of participants to the use of the interactions. The need to repeat the experiment several times or to deploy these technologies in living spaces is necessary.

Regarding the results obtained during the use case script, some limitations were highlighted, such as the slow response time of the controlled equipment, the poor performance of the voice recognition (poor response formulation, difficulties in opening the recognition channel... [6]) and the reliability of the system with the Internet network. There is also a strong need for visual or audio feedback according to the participant's profile on the action performed on the touch. The results of the logs [7] show successions of touch pointing for which it will be necessary to identify if they are due to motor inabilities or to a response time of the home automation objects.

The question "*which interface do you prefer to use according to your abilities*" shows that all 7 participants would like to alternate between a tactile and oral modality. 5 would prefer voice interaction and 2 (101 and 104) would prefer touch interaction. Participant 104 suggested physical interaction buttons. The possibility to have a choice of interaction based on one's abilities, environmental choice "verbatim: "*I find it very useful in certain contexts*", but also one's state (fatigue) is essential. The possibility to offer a multimodal interaction is also a request, because multimodality is considered in some cases as more efficient and faster. The low score of "novelty" dimension can be explained by the basic home automation controls for well-being.

## 5     Conclusion

Tactile and vocal commands gave the participants the possibility to control the equipment of the Smart Home. Thanks to the realization of controlled and free scenarios, the participants were able to live an experience that was observed by researchers. The participants were able to give their opinion, through different questionnaires. In terms of technological wishes, these analyses highlight the need to have a home automation control system that allows people with disabilities to control the different equipment in the house (windows, television, shutters, furniture, lights, etc.). This control must be able to centralize the functions of the house in order to avoid the multiplication of commands, it must also be customizable, mobile, understandable, easy to use and with several modes of interaction (touch or/and voice). However, these preliminary results must be put into perspective. Indeed, the people who carried out the tests were all technology-friendly and experienced.

## Acknowledgment

The study is partially funded by the Occitanie Region (France). The authors thank the participants and the GIHP association.